\documentclass[a4paper,11pt]{article}
\usepackage{pos}

\title{Full off-shell NLO QCD predictions for $t\bar{t}b\bar{b}$ at the LHC}

\author[a]{Manfred Kraus}

\affiliation[a]{Physics Department, Florida State University,\\
Tallahassee, FL 32306-4350, USA}

\emailAdd{mkraus@hep.fsu.edu}

\abstract{
We report on state-of-the-art theoretical predictions for $pp\to
t\bar{t}b\bar{b}$ including full off-shell effects in the di-lepton decay
channel at NLO QCD accuracy. We briefly discuss the impact of NLO QCD
corrections on differential distributions and also asses theoretical
uncertainties from scale and PDF dependence. Furthermore, we propose a simple
kinematic reconstruction in order to distinguish $b$ jets originating from
top-quark decays from arising from gluon splittings.
}

\FullConference{%
  The Tenth Annual Conference on Large Hadron Collider Physics - LHCP2022\\
  16-20 May 2022\\
  online
}

\begin{document}
\maketitle

\section{Introduction}
The precise understanding of the production of top-quark pairs in association
with $b$ jets is critical for current LHC and future HL-LHC measurements.  This
is especially true for the $pp\to t\bar{t}H$ production in the $H\to b\bar{b}$
decay channel, since $t\bar{t}b\bar{b}$ is the dominant QCD background and
therefore has direct impact on the top-quark Yuakawa measurements in this
channel. Furthermore, the $t\bar{t}b\bar{b}$ process is a main background for
the production of four top quarks, $pp\to t\bar{t}t\bar{t}$. Besides, being a
background to rare processes the $pp\to t\bar{t}b\bar{b}$ process is also
interesting by itself as a probe of QCD dynamics within a multi-scale
environment.

On the theoretical side the $pp\to t\bar{t}b\bar{b}$ process has been studied
in great detail. For instance, theoretical predictions at NLO QCD accuracy for
the on-shell production have been reported in Refs.~\cite{Bredenstein:2008zb,
Bredenstein:2009aj,Bevilacqua:2009zn,Bredenstein:2010rs,Worek:2011rd}.
Furthermore, $t\bar{t}b\bar{b}/t\bar{t}jj$ cross section ratios have been
studied in Ref.~\cite{Bevilacqua:2014qfa}.  Theoretical predictions for
$t\bar{t}b\bar{b}$ have also been matched to parton showers in the
$4$-flavor~\cite{Cascioli:2013era,Bevilacqua:2017cru,Jezo:2018yaf} as well as
in the $5$-flavor scheme~\cite{Kardos:2013vxa,Garzelli:2014aba}. To further
improve the understanding of QCD radiation patterns in this process also the
$pp\to t\bar{t}b\bar{b}j$ process has been studied at fixed-order in NLO QCD in
Ref.~\cite{Buccioni:2019plc}.  Eventually, first NLO predictions including full
off-shell effects for the $t\bar{t}b\bar{b}$ process in the di-lepton decay
channel have become available in Refs.~\cite{Bevilacqua:2021cit,Denner:2020orv}
as well as in the Narrow-width-approximation in Ref.~\cite{Bevilacqua:2022twl}.

Here, we report on our recent studies of the $pp\to t\bar{t}b\bar{b}$ process
\cite{Bevilacqua:2021cit,Bevilacqua:2022twl} in the di-lepton channel.

\section{Outline of the calculation}
We briefly summarize the outline of the calculations. For a detailed
description we refer the reader to Refs.~\cite{Bevilacqua:2021cit,
Bevilacqua:2022twl}. We compute NLO QCD corrections for $pp\to
b\bar{b}b\bar{b}e^+\nu_e\mu^-\bar{\nu}_\mu$ at
$\mathcal{O}(\alpha_s^5\alpha^4)$ for the LHC at $\sqrt{s}=13$ TeV.
\begin{figure}[h]
 \includegraphics[width=\textwidth]{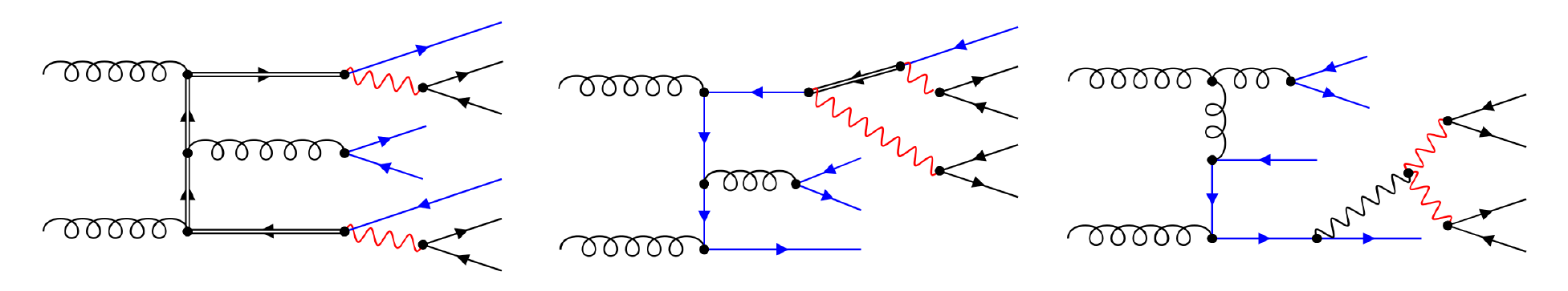}
 \caption{Illustrative Feynman diagrams for double, single and non-resonant
contributions to the full matrix elements of $pp\to
b\bar{b}b\bar{b}e^+\nu_e\mu^-\bar{\nu}_\mu$. Figure taken from
Ref.~\cite{Bevilacqua:2021cit}.}
 \label{fig:resonants}
\end{figure}
In this calculation, we employ matrix elements at the fully decayed stage.
Therefore, they include Feynman diagrams with double, single and non-resonant
top-quarks as shown in Fig.~\ref{fig:resonants}. The computation is performed
with the \textsc{Helac-Nlo}~\cite{Bevilacqua:2011xh} framework, that consists
out of \textsc{Helac-1Loop}~\cite{Ossola:2007ax,vanHameren:2009dr,
vanHameren:2010cp} for the evaluation of high-multiplicity one-loop matrix
elements and \textsc{Helac-Dipoles}~\cite{Czakon:2009ss,Bevilacqua:2013iha,
Czakon:2015cla} that deals with the infrared subtraction of the real radiation
contributions. The framework has been applied already for many $pp\to
t\bar{t}V$ processes including full off-shell effects to obtain
state-of-the-art predictions, see for instance Refs.~\cite{Bevilacqua:2015qha,
Bevilacqua:2016jfk,Bevilacqua:2018woc,Bevilacqua:2019cvp,Bevilacqua:2020pzy,
Bevilacqua:2021tzp,Bevilacqua:2022nrm}. Following ideas of
Ref.~\cite{Bern:2013zja} we store our theoretical predictions in form of
modified Les Houches Event files~\cite{Alwall:2006yp} and ROOT
Ntuples~\cite{Antcheva:2009zz}. This allows us to obtain predictions for
different scale and PDF choices via reweighting as well as to study new
observables.

\section{Phenomenological results}
In the following, we present results for for $pp\to b\bar{b}b\bar{b}
e^+\nu_e\mu^-\bar{\nu}_\mu$ at $\sqrt{s}=13$ TeV. We require two charged
leptons and at least four $b$ jets, where jets are formed using the anti-$k_T$
jet algorithm with $R=0.4$. Furthermore, we employ the following selection
cuts:
\begin{equation}
 p_T(\ell) > 20~\text{GeV}\;, \qquad p_T(b) > 25~\text{GeV}\;, \qquad
 |y(\ell)| < 2.5\;, \qquad |y(b)| < 2.5\;.
\end{equation}

\begin{figure}[h]
 \includegraphics[width=0.48\textwidth]{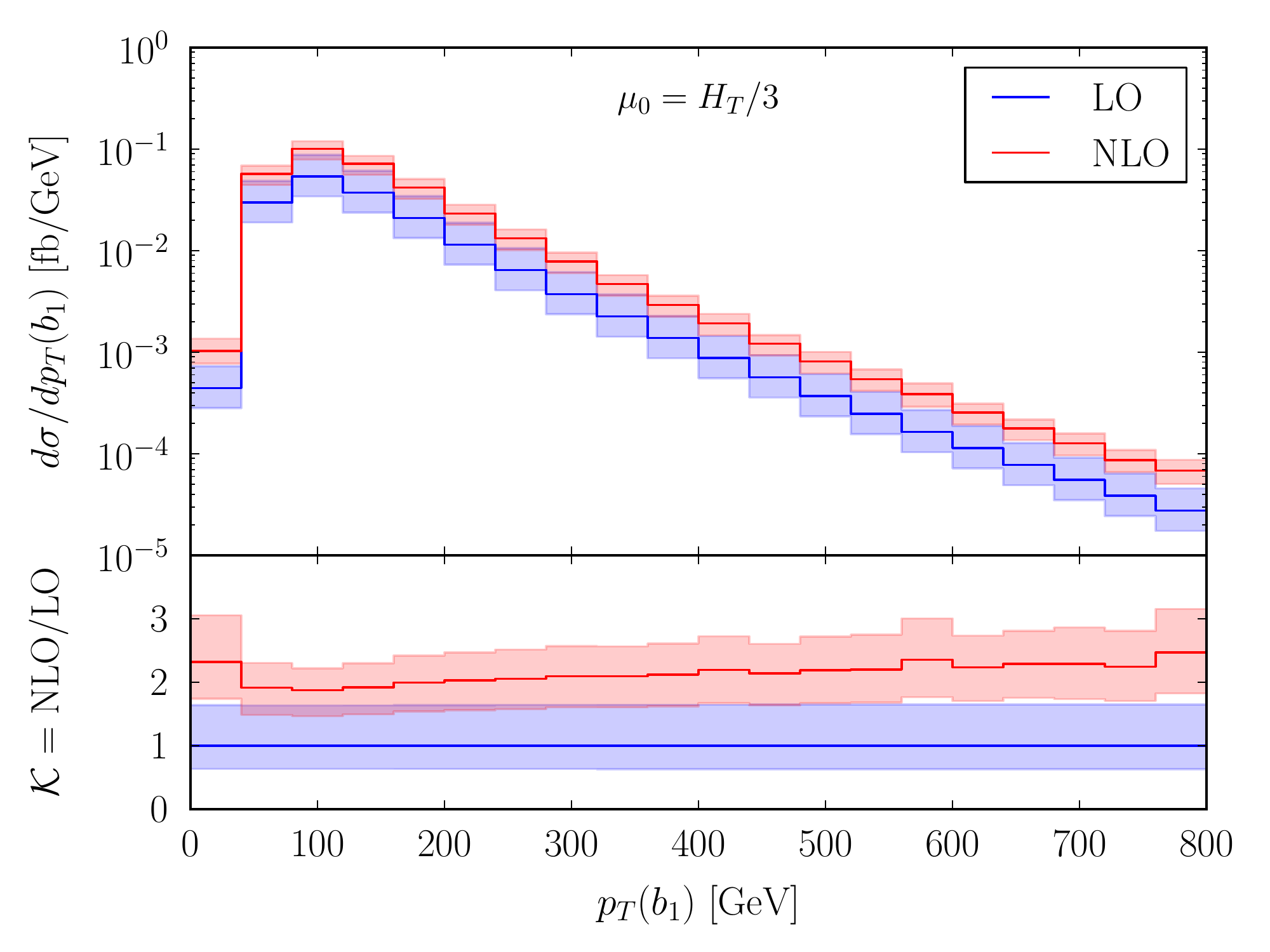}
 \includegraphics[width=0.48\textwidth]{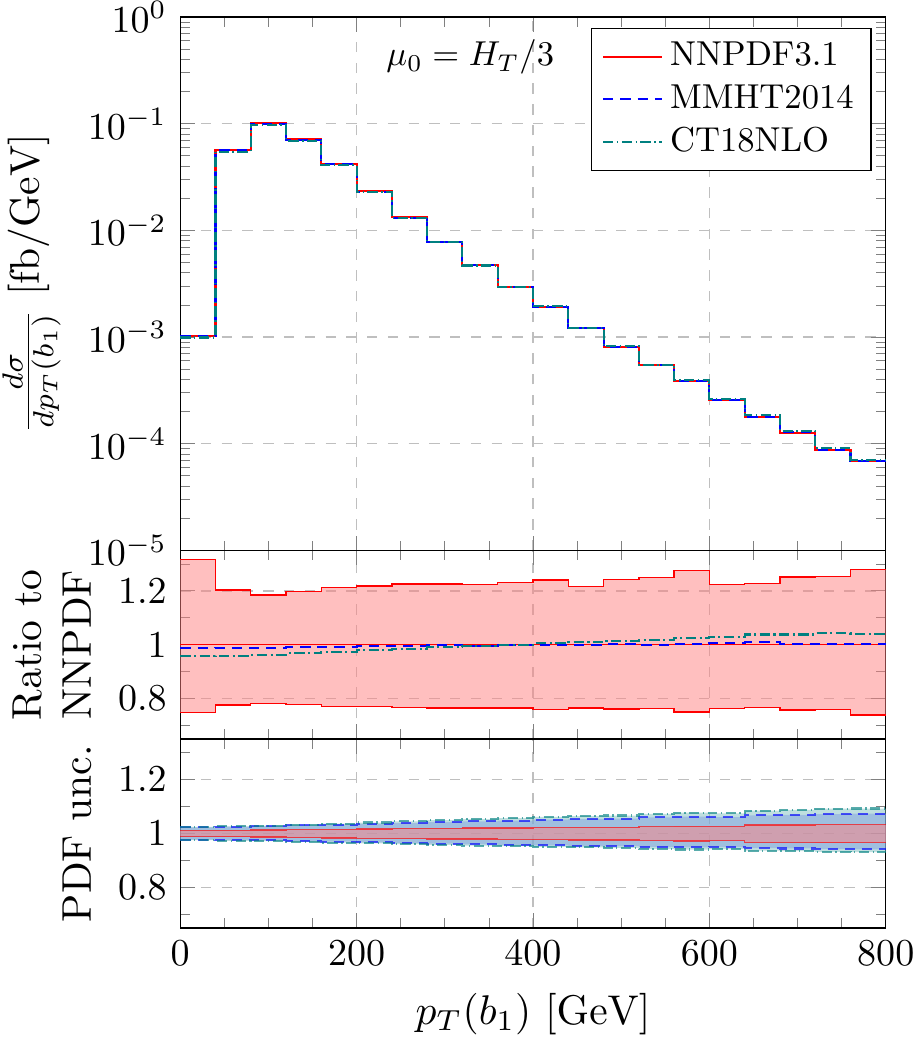}
 \caption{Theoretical predictions for the transverse momentum of the leading
$b$ jet. Figures taken from Ref.~\cite{Bevilacqua:2021cit}.}
 \label{fig:res_1}
\end{figure}
In Fig.~\ref{fig:res_1} the transverse momentum of the leading $b$ jet is
shown.  The plot on the left hand-side shows the impact of NLO QCD corrections
on the spectrum.  We observe that NLO QCD corrections are large and of the
order of $90-135\%$.  Furthermore, the theoretical uncertainties, estimated by
independent variations of the renormalization and factorization scales, are
reduced to $20-30\%$ at NLO. Even though the uncertainties are still sizable
they barely overlap with the uncertainty estimates at leading order. On the
right hand-side of Fig.~\ref{fig:res_1} we also show PDF uncertainties, which
are at most of the order of $\pm 10\%$. Therefore, theoretical uncertainties
are still dominated by missing higher-order corrections.

\begin{figure}[h]
 \includegraphics[width=0.48\textwidth]{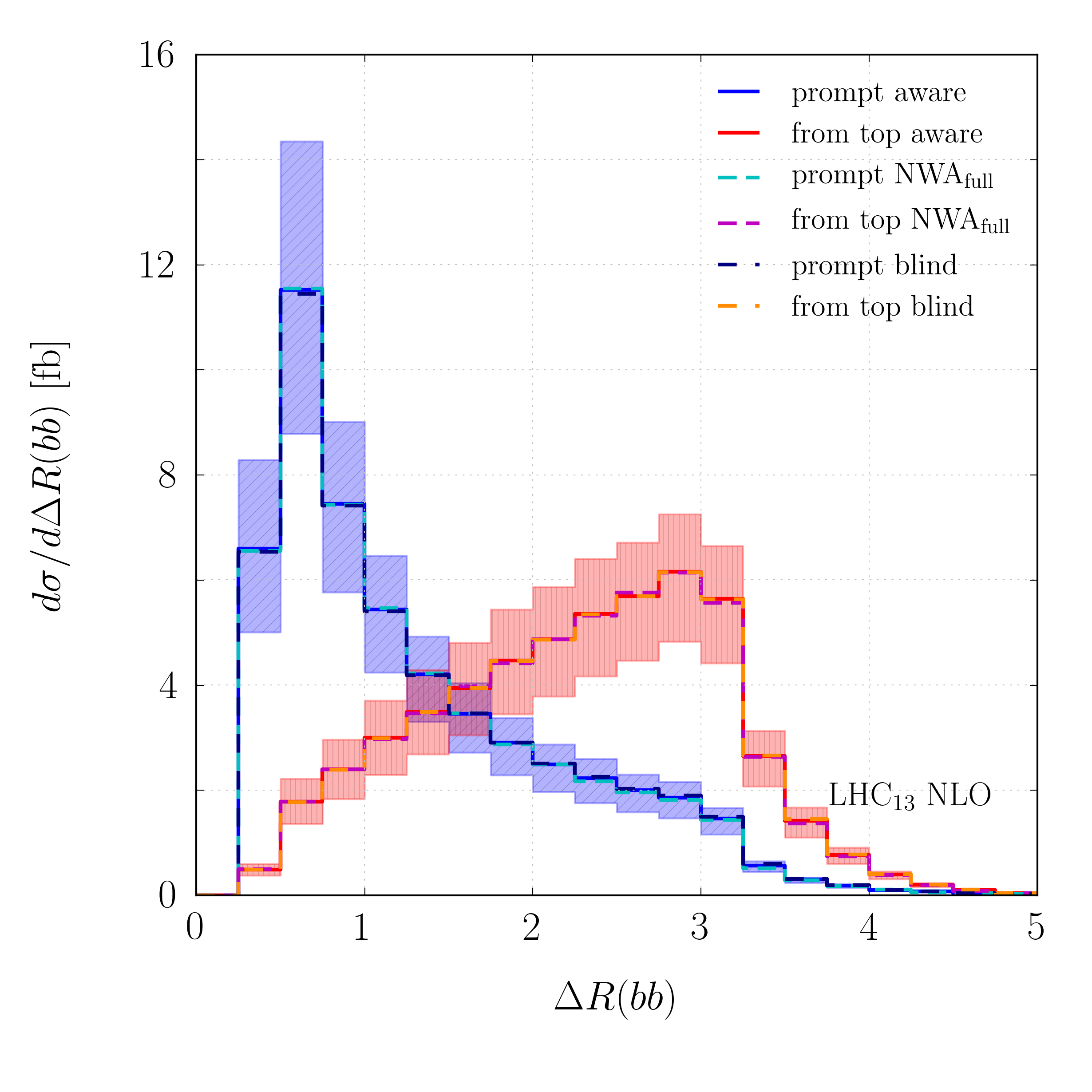}
 \includegraphics[width=0.48\textwidth]{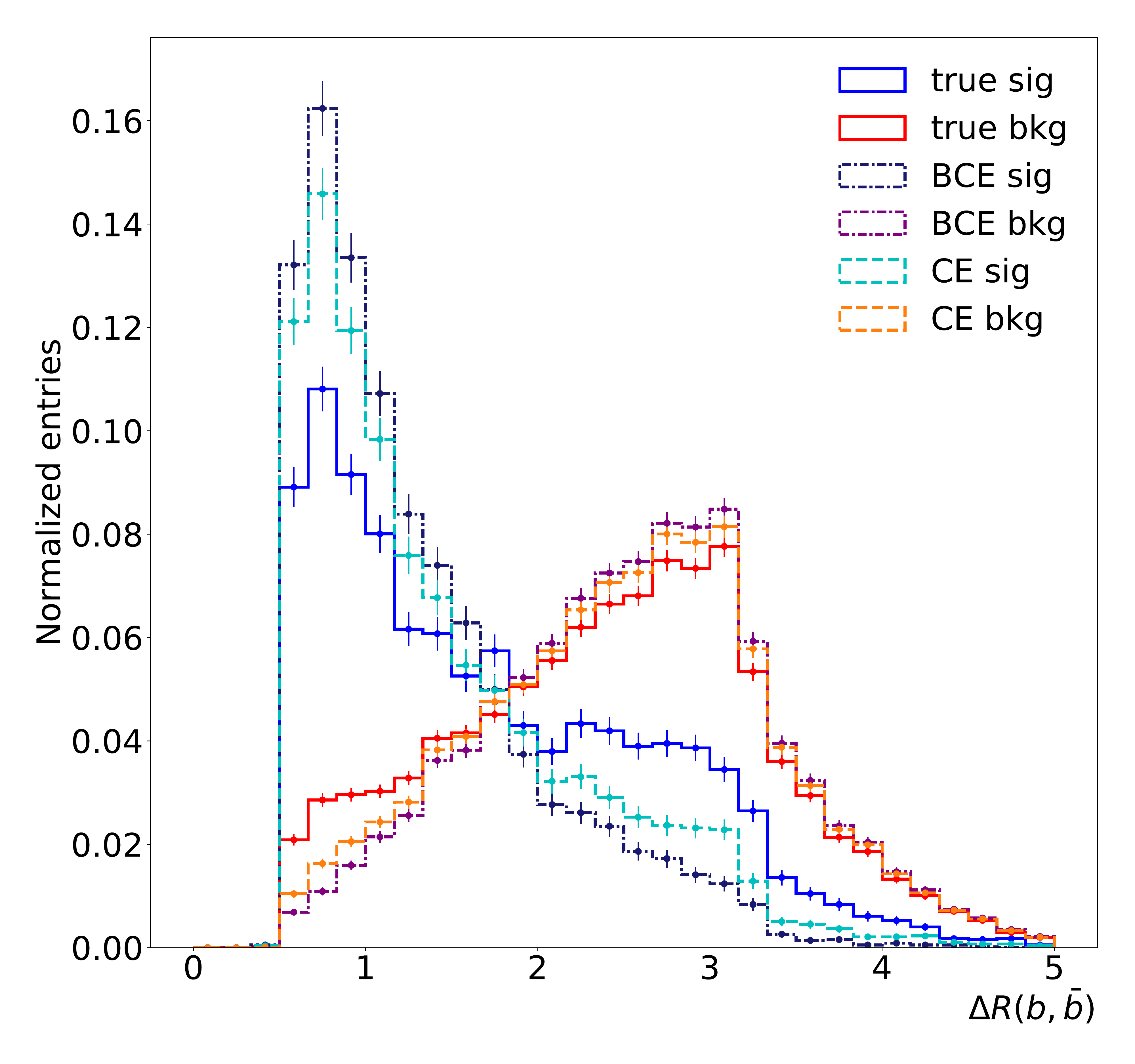}
 \caption{$\Delta R(bb)$ distribution for prompt $b$ jets and those of top
decays. Left figure taken from Ref.~\cite{Bevilacqua:2021cit} and right figure
from Ref.~\cite{Jang:2021eph}.}
 \label{fig:res_2}
\end{figure}
We also studied the possibility to differentiate between $b$ jets originating
from top decays and \textit{prompt} $b$ jets stemming from $g\to b\bar{b}$
splittings. Being able to distinguish these jets could be beneficial for 
$pp\to t\bar{t}H$ measurements in the $H\to b\bar{b}$ decay channel. We found
that, by minimizing the following function 
\begin{equation}
 Q = \left| M(t) - m_t \right| \times \left| M(\bar{t}) - m_t\right| 
 \times M^{\textrm{prompt}}(bb)
\end{equation}
over all possible momentum assignments for $t,\bar{t}$ and $bb$, we achieve a
good categorization of the $b$ jets. For instance, on the left side of
Fig.~\ref{fig:res_2} the $\Delta R(bb)$ observable for prompt $b$ jets and
those from top decays is shown using our method. Our simple kinematic
reconstruction is consistent with results obtained by employing Neural
Networks~\cite{Jang:2021eph} as shown on the right of
Fig.~\ref{fig:res_2}.

\section{Summary}
We presented some of our recent results for full off-shell $t\bar{t}b\bar{b}$
production at the LHC. We have investigated in detail the impact of NLO QCD
corrections on differential distributions. Furthermore, we estimated
theoretical uncertainties by scale variations and the incomplete knowledge of
PDFs. We found large NLO corrections of the order of $90\%$ and the theoretical
uncertainties are clearly dominated by missing higher-order corrections.
In addition, we proposed a simple kinematic reconstruction that allows us to 
distinguish $b$ jets from top-quark decays and prompt $b$ jets stemming from
$g\to b\bar{b}$ splittings. For more detailed information we refer to the
reader to Refs.~\cite{Bevilacqua:2021cit,Bevilacqua:2022twl}.

\begin{description}
 \item[Acknowledgements:] This work is supported in part by the U.S. Department
 of Energy under the grand DE-SC010102.
\end{description}

\end{document}